\documentclass[letter]{aa} 

\usepackage{lscape,graphicx,natbib,moreverb,verbatim,fancyvrb}
\usepackage{txfonts}

\newcommand{\xhi}{x_\mathrm{HI}}
\newcommand{\nhi}{N_\mathrm{HI}}

\begin{document}

   \title{The reanalysis of spectra of GRB 080913 to estimate the neutral fraction of the IGM at a redshift of 6.7}

   \author{M. Patel
     \inst{1}
     \and
     S. J. Warren
     \inst{1}
     \and 
     D. J. Mortlock
     \inst{1}
     \and
     J. P. U. Fynbo
     \inst{2}
   }

   \institute{Astrophysics Group, Imperial College London, Blackett Laboratory, Prince Consort Road, London, SW7 2AZ, United Kingdom\\
              \email{m.patel06@imperial.ac.uk}
         \and
             Dark Cosmology Centre, Niels Bohr Institute, University of Copenhagen, Juliane Maries Vej 30, DK-21000, Copenhagen, Denmark\\
             }

   \date{Received ; accepted }

\abstract{} {We reanalyse optical spectra of the $z=6.7$ gamma-ray
  burst GRB 080913, adding hitherto unpublished spectra, in order to
  reassess the measurement of the neutral fraction of the IGM at high
  redshifts.}  {In the data reduction we take particular care to
  minimise systematic errors in the sky subtraction, which are evident
  in the published spectrum, and compromise the analysis.  The final
  combined spectrum has a higher S/N than the previously published
  spectrum by a factor of 1.3.}  {We find a single significant
  absorption line redward of the Ly$\alpha$ continuum break, which we
  identify with the SII+SiII $\lambda 0.126\mu$m blend, at
  $z=6.733$. The sharp spectral break at Ly$\alpha$ implies a
  comparatively low total column density of neutral hydrogen along the
  line of sight, $\log{(\nhi/\mathrm{cm^{-2}})}<20$. We model the absorption with a
  host-galaxy DLA, surrounded by an ionised region of unknown size
  $r$, within the IGM of neutral fraction, $\xhi$. Despite knowing the
  source redshift, and the improved S/N of the spectrum, when
  fitting only over wavelengths redward of Ly$\alpha$,
  no useful constraints on $\xhi$ can be obtained. We consider the
  possibility of including the ionised region, blueward of Ly$\alpha$,
  in constraining the fit. For the optimistic assumption that the
  ionised region is transparent, $\tau_{GP}\ll 1$, we find that the
  region is of small size $r<2\,$ proper Mpc, and we obtain an upper
  limit to the neutral fraction of the IGM at $z=6.7$ of $\xhi<0.73$
  at a probability of 90$\%$.}  {}
 

\maketitle

\section{Introduction}

The hydrogen in the Universe became predominantly neutral at the epoch
of recombination, $z=1070$. The inter-galactic medium (IGM) out to
$z=5$, however, is highly ionised. Two observations made within the
last decade have narrowed the window within which cosmic hydrogen
reionisation took place. First, the electron scattering optical depth
to the cosmic microwave background measured by the \textit{Wilkinson
Microwave Anisotropy Probe} implies a redshift of reionisation of
$11.4\pm 1.4$ \citep{Dunkley_etal:2009}. Second, observations of the
Ly$\alpha$ forest in the spectra of $z\sim6$ quasars indicate that
$z=5.8$ marks the tail-end of the epoch of reionisation
\citep{Fan_etal:2006}. These two results suggest that reionisation is
an extended process.  Therefore, in order to understand the chronology
of reionisation in detail, there is considerable interest in detecting
sources beyond $z=6.4$, the redshift of the most distant quasars so
far discovered \citep{Fan:2003, Willott:2007}.

Ly$\alpha$ emitting galaxies (LAEs) have been detected out to $z=7$
\citep{Kashikawa:2006, Ota_etal:2008}. Since the strength of the
Ly$\alpha$ emission line depends on the neutral fraction, $\xhi$, one
approach advocated to chart the progress of reionisation
\citep{McQuinn:2007} is to measure the evolution of the properties
(e.g. abundance, clustering, etc) of LAEs. However, the processes
involved are difficult to model accurately
\citep[e.g.][]{Tasitsiomi:2006, Dijkstra:2007}, and there is little
consensus yet on the interpretation of the results from this approach.

The confirmation of the cosmological nature of gamma ray bursts (GRBs)
\citep{Kulkarni_etal:1998} revealed their potential as probes of the
high-redshift Universe. \cite{Barkana:2004} emphasise two particular
advantages of GRBs over quasars for studying reionisaton: that their
afterglows will be visible to redshifts of $z\sim10$; and that
the size of the ionised region around the host galaxy, which
complicates the interpretation of the spectrum, will be small. Recent
numerical simulations by \cite{McQuinn_etal:2008} and
\cite{Mesinger_Furlanetto:2008}, however, suggest that GRB host
galaxies may be located in dense environments, and therefore the IGM
immediately surrounding the host galaxy may be ionised by nearby
quasars or local massive star forming galaxies, and so the size
of the ionised region needs to be included in the modelling. In any
case a high signal-to-noise ratio (S/N) spectrum is needed to
distinguish between the different signatures of a high-column density
of neutral hydrogen in the GRB host galaxy (or nearby)
\citep{RuizVelasco_etal:2007}, and of distributed neutral hydrogen in
the IGM.

Until the launch of the {\em Swift} satellite
\citep{Gehrels_etal:2004}, the furthest known GRB was at $z=4.5$
\citep{Andersen_etal:2000}, compared to $z=5.8$ for the furthest known
quasar at that time \citep{Fan_etal:2000}. Since then, three GRBs at
$z>6$ have been discovered using {\em Swift}: GRB 050904 at $z=6.3$ in
2005 \citep{Cusumano_etal:2006, Kawai_etal:2006}; GRB 080913 at
$z=6.7$ in 2008 \citep{Greiner_etal:2009}; and, in 2009, the
remarkable source GRB 090423 at $z=8.2$
\citep{Tanvir_etal:2009,Salvaterra_etal:2009}, the most distant object
yet found.

\cite{Totani_etal:2006} present a detailed analysis of the optical
afterglow spectrum of GRB 050904, which was measured with the highest
S/N of the three $z>6$ GRBs. The spectrum displays a red damping wing
from Ly$\alpha$ absorption that is produced by some combination of a
high-column density absorber near the GRB (hereafter referred to as a
DLA, for damped Ly$\alpha$ absorber), and a smoothly distributed
component in the IGM. By fitting absorption models containing these
two components, \cite{Totani_etal:2006} were able to place a limit on
the neutral fraction of the IGM at $z=6.3$ of $\xhi<0.17(0.60)$ at
$68\%(95\%)$ confidence.
The constraints are relatively weak despite the reasonably high S/N of
the spectrum. The sources GRB 080913 and GRB 090423 are potentially
more interesting, because of their higher redshifts, but the
published spectra have insufficient S/N to place any useful
constraints on $\xhi$ when employing two-component (DLA+IGM) fits.

Here we describe and analyse an improved spectrum of GRB 080913, which
includes unpublished spectroscopic data taken three nights after the
published spectrum. In Section 2 we describe the observations taken on
each night, and the reduction techniques. We analyse the combined
spectrum in Section 3, first searching for absorption lines in the new
spectrum, in order to measure the redshift of the source, and then
fitting a two-component DLA+IGM model to the observed continuum
break. Finally, the results are summarised in Section 4.

\section{Observations and Data reduction}

\cite{Greiner_etal:2009} (hereafter Paper I) provide a summary of all
the photometric and spectroscopic observations of GRB 080913. Here we
recap the details of the spectroscopic observations only. The object
was first observed at the Very Large Telescope using the FOcal
Reduction and low dispersion Spectrograph 2 instrument on the night
beginning Septempber 13 2008. A 1 arcsec slit and the 600z grism were
employed, providing wavelength coverage from 0.7470 to
$1.0570\mu$m. One 1800s and one 600s exposure were taken, but the
second exposure is not useful as it was taken in twilight. The first
spectrum is published in Paper I, and used for the analysis presented
there. Due to poor weather, further observations were not possible
until three nights later, on the night beginning September 16 2008,
when 7 exposures of 1800s each were secured. Because the source was
then fainter, the S/N of the combined spectrum from the second night
is lower than that of the single spectrum from the first night.

We reduced the data in the following manner. After the standard bias
subtraction and flat-fielding steps, cosmic rays were removed from
individual frames using the Laplacian Cosmic Ray Removal Algorithm of
\cite{vanDokkum:2001}. The 7 frames observed on the second night were
taken using 4 different slit positions. We adapted methods from
near-infrared spectroscopy described in \cite{Weatherley:2005} to
minimise residuals from the subtraction of bright sky lines. After
subtracting a functional fit up each column in every frame
(first-order sky subtraction), all frames at other slit positions were
averaged and subtracted (second-order sky subtraction). The 7 frames
were then registered and combined, weighting by the inverse
variance. Since only one frame was taken on the first night, frames
from the second night were used for the second-order sky subtraction.
Compared to the standard method for optical long-slit spectroscopy
(i.e. first-order sky subtraction), the method followed here results
in a slight increase in the random errors, but largely eliminates
systematic errors.

One-dimensional optimal spectral extraction was performed separately
for each night, weighting by the profile of a nearby bright
star. Wavelength calibration was achieved using observations of a
HeNeAr lamp. Corrections for telluric absorption and flux calibration
were applied simultaneously using observations of the standard star
LTT 7987. We scaled the spectrum from the second night to the spectrum
from the first night, and combined the data, weighting by the inverse
variance. The flux calibration agrees well with that in Paper I.

\section{Analysis of the Optical Afterglow Spectrum}

\begin{figure*}
\centering
\includegraphics[width=75mm, height=130mm, angle=90]{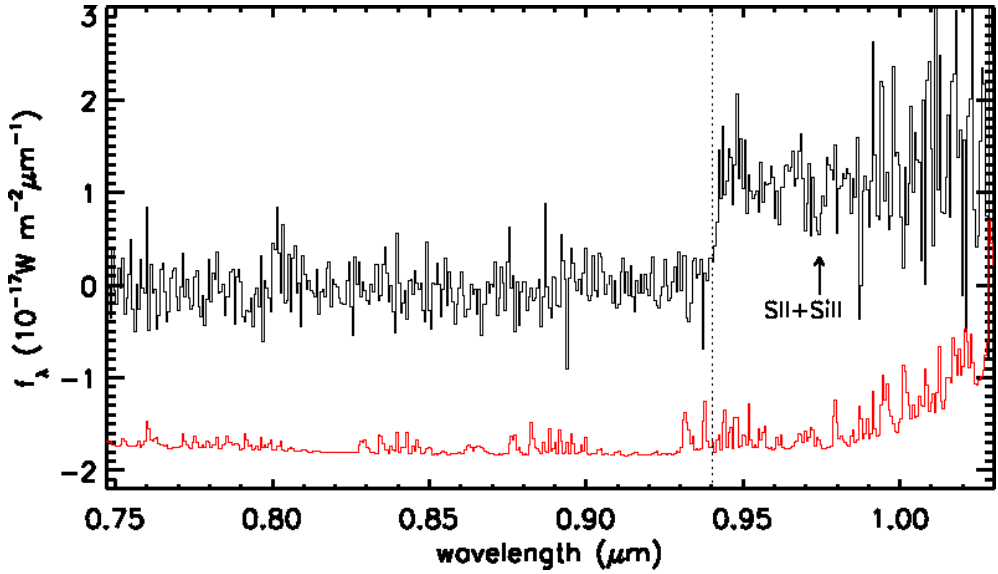}
\caption{Combined VLT/FORS2 spectrum of GRB 080913. The data have been
  binned by a factor 4, providing 2 binned pixels per resolution
  element. The $1\sigma$ error spectrum is plotted in red and is
  offset by $-2.0\times10^{-17}$ Wm$^{-2}\mu$m$^{-1}$ for clarity. The
  position of the SII+SiII absorption line (as indicated), implies a
  source redshift of 6.733. The dotted line corresponds to redshifted
  Ly$\alpha$.}
\label{F:VLT_combined}
\end{figure*}

The final combined spectrum, binned to critical sampling, is shown in
Fig. \ref{F:VLT_combined}. The S/N of the new spectrum compared to the
original spectrum shows an improvement by a factor of 1.3 on average. The
sharp break shows no clear evidence of a significant red damping wing,
and is indicative of a relatively low column density of neutral gas,
compared to GRB 050904.

\subsection{Search for absorption lines}
The Ly$\alpha$ continuum break near $0.94\mu$m corresponds to
$z\sim6.73$. In the hope of measuring an accurate redshift, we made a
search for absorption lines in the afterglow spectrum.

For all the fits presented in this paper we assumed that the spectral
energy distribution of the GRB follows a power-law
($F_{\nu}\propto\nu^{-\alpha}$), where $\alpha=1.12$, which was
derived from photometric measurements taken on the first night (Paper
I). After subtracting the continuum fit, we tested for the presence of
an absorption line centred at each pixel redward of the break, and
then refined the measurement of the line centre of any detected lines,
by refitting with the central wavelength as a free parameter. We
assumed a Gaussian profile matched to the spectral resolution (i.e. 8
unbinned pixels, FWHM $=0.00128\mu$m). The S/N at any pixel $j$ is
given by:
\begin{equation}
(S/N)_{j} = \frac{\sum f^{\,\mathrm{r}}_{j+i}u_{i}/\sigma^{2}_{j+i}}{\sqrt{\sum u_{i}^{2}/\sigma^{2}_{j+i}}},
\end{equation}
where $u$ describes the line profile, centred on $i=0$,
$f^{\,\mathrm{r}}$ is the residual flux, and $\sigma$ is the flux
error \citep{Bolton_etal:2004}. 

Over the rest-frame wavelength range measured, one of the strongest
absorption lines typically seen in afterglow spectra is the SII+SiII
$0.1260\mu$m blend, which is expected to be situated near
$0.974\mu$m. We found only one absorption line with significance
greater than $S/N=2$, detected at the wavelength $0.9743\mu$m at
$S/N=2.9$. The line is marked in Fig. \ref{F:VLT_combined}. The
wavelength matches the expected location of the SII+SiII $0.1260\mu$m
line, which is unlikely to be a coincidence. Therefore we infer that
the line is real. The measured wavelength provides an absorption
redshift of $z_{\mathrm{abs}}=6.733$.

\subsection{Fitting the continuum break}

The continuum break is due to a combination of absorption by neutral
gas in the GRB host galaxy and the IGM, therefore only a two-component
(DLA+IGM) fit provides a meaningful physical model. The modelling is
complicated by the unknown size and neutral fraction of the
surrounding ionised region. For this reason it is useful first to fit
single-component models to quantify the strength of the break.

For each single-component model, the power-law SED is modified by
absorption by either a DLA or the IGM, computed using the equations in
\cite{Totani_etal:2006}. We limit the fits to data blueward of
$1.00\mu$m, and mask out the SII+SiII absorption line. For the DLA
model we assume complete absorption blueward of the line centre, since
this is achieved for a minimal value of $\xhi\sim10^{-5}$. The fits
are determined using $\chi^{2}$ minimisation.

\begin{figure}
\centering
\includegraphics[width=55mm, height=90mm, angle=90]{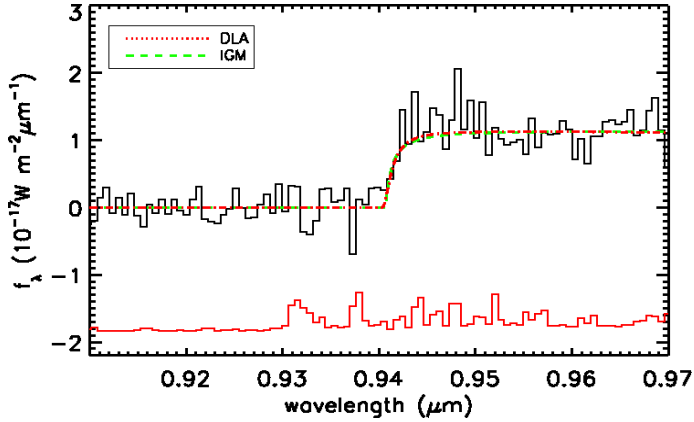}
\caption{Best-fit single-component DLA and IGM models, overplotted
on the GRB afterglow spectrum, which is rebinned by a factor of 4.}
\label{F:grb080913_bf}
\end{figure}

The DLA model has three free parameters: the redshift,
$z_{\mathrm{DLA}}$; the column density, $\nhi$; and the
continuum normalisation at a wavelength of $0.96\mu$m,
$c_{\mathrm{DLA}}$.
We find best-fit values of $\log{(\nhi/\mathrm{cm^{-2}})}=19.84$, and
$z_{\mathrm{DLA}}=6.731$. The redshift is in excellent agreement with
the absorption-line redshift.  The value of $\nhi$ is low in
comparison with the majority of GRB-DLA systems detected so far
\citep{Fynbo_etal:2009}, which is interesting as this is in line with
the prediction of \cite{Nagamine_etal:2008} that the typical column
densities of GRB-DLA systems decrease towards higher redshifts.  The
IGM model is defined by four parameters: the neutral fraction $\xhi$,
the continuum normalisation at $0.96\mu$m, $c_{\mathrm{IGM}}$, and the
upper and lower redshift limits over which $\xhi$ applies,
$z_{\mathrm{IGM, u}}$ and $z_{\mathrm{IGM, l}}$. We fix the lower
redshift to z$_{\mathrm{IGM, l}}=6.0$.
We find best-fit values of $\xhi = 0.21$, and $z_{\mathrm{IGM,
u}}=6.737$. The two models are plotted in
Fig. \ref{F:grb080913_bf} and are almost
indistinguishable, implying that there will be a strong degeneracy
between the parameters $\nhi$ and $\xhi$ in a two-component fit.

These two fits are quite different from those obtained by
\cite{Greiner_etal:2009}, who found much stronger absorption. For the
DLA model their $95.4\%$ confidence range is
$20.29<\log{(\nhi/\mathrm{cm^{-2}})}<21.41$, and for the IGM model
they found a best fit value $\xhi=1.00$, with $\xhi>0.35$ at $95.4\%$
confidence.
From an analysis of the two spectra, the differences may be explained
by systematic sky-subtraction residuals in their spectrum, caused by
the strong OH sky lines in the region $0.93-0.96\mu$m (visible in the
error spectrum, Fig. \ref{F:VLT_combined}). We demonstrate this by a
quantitative comparison of the two spectra and error arrays. We first
binned each spectrum by a factor of four, summing the variance arrays
appropriately. We then subtracted a median-filtered version of the
data, and divided the result by the error array. In the absence of
systematic errors, or significant features in the spectra, these S/N
spectra should have mean zero and $\sigma_{\mathrm{S/N}}\sim1.0$.  We
find this to be true for our spectrum, at all wavelengths, and also
for the spectrum of \cite{Greiner_etal:2009} in regions free of strong
sky lines. However over the wavelength range of interest
$0.93-0.96\mu$m we measure $\sigma_{\mathrm{S/N}}=1.7$ in their
spectrum, which is strong evidence of systematic sky-subtraction
residuals.

Turning now to the two-component model, we firstly fix
$z_{\mathrm{DLA}}$ to the absorption redshift of $z=6.733$, since the
absorption presumably arises in the dominant neutral-gas system in the
host galaxy. However, we must consider the possibility that
$z_{\mathrm{DLA}}$ and $z_{\mathrm{IGM, u}}$ differ, because the GRB
host galaxy is surrounded by an ionised region. The two-component
model therefore requires five parameters: $\xhi$, $z_{\mathrm{IGM,
    u}}$, $z_{\mathrm{DLA}}$, $\nhi$ and the continuum normalisation
at $0.96\mu$m, $c$. A further complication is the uncertain degree of
absorption within the ionised bubble. The GP
\citep{Gunn_Peterson:1965} optical depth is given by
\begin{equation}
\tau_{GP}=5.4\left(\frac{1+z}{7.7}\right)^{3/2}\left(\frac{\xhi}{10^{-5}}\right)(1+\Delta),
\end{equation}
where $\Delta$ is the overdensity. If $\tau_{GP}$ is large, there is
negligible transmission blueward of Ly$\alpha$, so that only the
wavelength range redward of Ly$\alpha$ is useful. Under these
conditions, at this low S/N, we find we can obtain no useful
constraints on the various parameters of the two-component model
\---\ for the IGM component, which only contributes significantly to
the wing of the absorption profile, a large ionised region and large
$\xhi$ provides a similarly good fit as a small ionised region and
small $\xhi$. On the other hand, if $\tau_{GP}$ within the ionised
region is small, then, for a large ionised region, there will be
measureable transmission to the blue of the DLA line centre: see
\citet{Iliev_etal:2008} for illustrations of such cases from
simulations. Under these conditions, with high S/N, this region could
also be used in the modelling. For our spectrum, due to the low S/N,
we consider only the simple (optimistic) case that the ionised
region is completely transparent, $\tau_{GP}\ll 1$. We assume a
uniform prior on each parameter except for the column density. Since
GRBs are most likely to occur in regions with significant neutral
hydrogen, we apply a prior which scales linearly with column density.

\begin{figure}
\centering
\includegraphics[width=70mm, height=60mm]{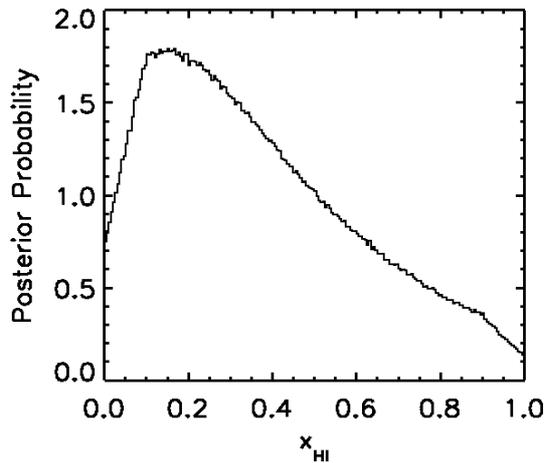}
\caption{The posterior probability distribution of the neutral fraction when a joint DLA$+$IGM fit is employed. All other parameters have been marginalized.}
\label{F:grb080913_xhi_prob}
\end{figure}

\begin{figure}
\centering
\includegraphics[width=70mm, height=60mm]{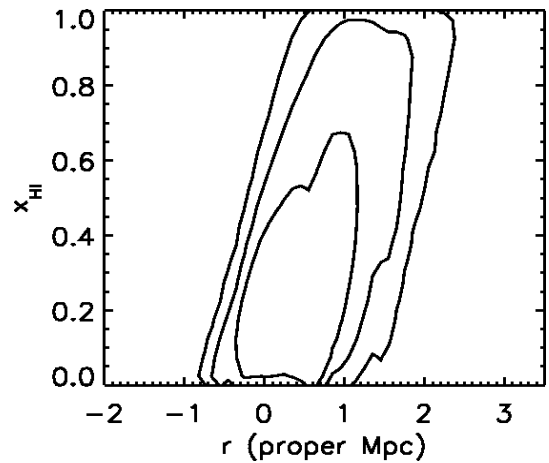}
\caption{Posterior probability contours of the joint DLA+IGM fit in $r$-$\xhi$ space. All other parameters have been marginalized. Regions enclosing $68.2\%$, $95.4\%$ and $99.7\%$ of the posterior probability are highlighted.}
\label{F:grb080913_contour_r_xhi}
\end{figure}

We find best-fit parameters of $\log{(\nhi/\mathrm{cm^{-2}})}=19.60$,
$\xhi=0.06$, $z_{\mathrm{IGM, u}}=6.733$. To determine the
uncertainties in each of these parameters we use a Markov Chain Monte
Carlo (MCMC) algorithm and sample the posterior probability
distribution. The result is presented in
Fig. $\ref{F:grb080913_xhi_prob}$, which shows the posterior
probability distribution of the IGM neutral fraction having
marginalised over all other parameters. We find $\xhi<0.73$ with a
probability of $90\%$. Our analysis also provides constraints on the
size of the ionised region around the GRB host.  In
Fig. $\ref{F:grb080913_contour_r_xhi}$, we plot the marginalized
posterior probability contours in $\xhi$-$r$ space produced from our
fits. Here we see that an ionised region which has a size smaller than
$\sim2\,$ proper Mpc is favoured. The regions in parameter space where
r is negative correspond to situations where the DLA material is
accelerated towards us due to either the GRB event or its
progenitor. If we ignore these regions, we find that $r<1.3$ proper
Mpc with a probability of $90\%$, which suggests that a large ionised
region is not present around the host galaxy. This value is, however,
still consistent with external sources of ionising flux being
present. For comparison \cite{Haiman:2002} found that the ionised
regions around LAEs have a typical size of $0.8\,$ proper Mpc.


\section{Conclusion}

New optical spectroscopic observations of GRB 080913 have been
presented and analysed. The detection of SII+SiII absorption
($0.1260\mu$m) at $2.9\sigma$ provides a redshift of the DLA host
galaxy of $z=6.733$. Employing a joint DLA$+$IGM model to fit the
observed continuum break, we find an upper limit to the neutral
fraction of the IGM $\xhi<0.73$ at a probability of $90\%$. However 
this result rests on the assumption that the
ionised region surrounding the host galaxy is transparent to Ly$\alpha$.
Any analysis of a GRB spectrum needs to include the
radius of the ionised region as a free parameter, and to consider the
question of the neutral fraction within this zone. 
Furthermore the
scatter in measurements of $\xhi$ between different sources at similar
redshifts is predicted to be substantial
\citep{McQuinn_etal:2008,Mesinger_Furlanetto:2008}, and needs to be
quantified.  Higher S/N spectra of several sources at high redshift
will be required to make significant progress in this field.

\begin{acknowledgements}
M.P. acknowledges funding from the University of London. The Dark
Cosmology Centre is funded by the DNRF. We are grateful to the referee
for comments which helped improve the manuscript substantially.
\end{acknowledgements}

\bibliography{grb080913_ref}
\bibliographystyle{aa}


\end{document}